\documentclass[referee,a4paper,12pt,traditabstract,dvipdfmx]{swsc} 

\usepackage{amsmath}
\usepackage{graphicx}
\usepackage{txfonts}
\usepackage{subfigure}
\usepackage{lineno}
\usepackage{url}
\usepackage{multirow}
\usepackage{arydshln}
\usepackage{color}

\begin{document}


   \title{Verification of operational solar flare forecast}

   \subtitle{Case of Regional Warning Center Japan}
   
   \titlerunning{Flare forecast verification}

	\authorrunning{Kubo, Den, \and Ishii}

   \author{Y\^uki Kubo,
		Mitsue Den,
		\and
		Mamoru Ishii
          }

   \institute{National Institute of Information and Communications Technology (NICT),
              Tokyo 184-8795, Japan\\
		\email{kubo@nict.go.jp}
             }


 
  \abstract
{In this article, we discuss a verification study of an operational solar flare forecast in the Regional Warning Center (RWC) Japan. The RWC Japan has been issuing four-categorical deterministic solar flare forecasts for a long time. In this forecast verification study, we used solar flare forecast data accumulated over 16 years (from 2000 to 2015). We compiled the forecast data together with solar flare data obtained with the Geostationary Operational Environmental Satellites (GOES). Using the compiled data sets, we estimated some conventional scalar verification measures with 95\% confidence intervals. We also estimated a multi-categorical scalar verification measure. These scalar verification measures were compared with those obtained by the persistence method and recurrence method. As solar activity varied during the 16 years, we also applied verification analyses to four subsets of forecast-observation pair data with different solar activity levels. We cannot conclude definitely that there are significant performance difference between the forecasts of RWC Japan and the persistence method, although a slightly significant difference is found for some event definitions. We propose to use a scalar verification measure to assess the judgment skill of the operational solar flare forecast. Finally, we propose a verification strategy for deterministic operational solar flare forecasting. For dichotomous forecast, a set of proposed verification measures is a frequency bias for bias, proportion correct and critical success index for accuracy, probability of detection for discrimination, false alarm ratio for reliability, Peirce skill score for forecast skill, and symmetric extremal dependence index for association. For multi-categorical forecast, we propose a set of verification measures as marginal distributions of forecast and observation for bias, proportion correct for accuracy, correlation coefficient and joint probability distribution for association, the likelihood distribution for discrimination, the calibration distribution for reliability and resolution, and the Gandin-Murphy-Gerrity score and judgment skill score for skill.}


   \maketitle

\section{Introduction}
\label{sec:intro}
Forecast verification has been recognized as one of the most important topics in terrestrial weather forecast research. In the field of space weather forecast, however, forecast verification is still in the early stages. Recently, some Regional Warning Centers (RWCs), belonging to the International Space Environment Service (ISES), have started to verify their operational forecasts. As a result of this initiative, forecast verification has become recognized as one of the important research topic in the operational space weather forecasting community.\par

RWC USA (Space Weather Prediction Center at the National Oceanic and Atmospheric Administration: SWPC/NOAA) verified their operational solar flare forecast by using forecast data accumulated over about 12.5 years (Crown 2012). RWC Belgium (Royal Observatory of Belgium: ROB) also verified their operational solar flare forecast and geomagnetic K-index forecast by using forecast data accumulated over about 8.5 years\footnote{They also verified their F10.7 forecast for data accumulated over 11 years.} (Devos et al. 2014). Other RWCs have also started verification studies of their operational space weather forecasts. \par

Recently, the Community Coordinated Modeling Center has been planning the Flare Scoreboard, which is an online platform of real-time probabilistic solar flare forecast verification (http://ccmc.gsfc.nasa.gov/challenges/flare.php). RWC Japan (National Institute of Information and Communications Technology: NICT) also has an on-line platform of operational solar flare and geomagnetic K-index forecast verification (http://seg-web.nict.go.jp/cgi-bin/forecast/eng\_forecast\_score.cgi), which compares forecasts of some RWCs. However, as the forecast conditions are not the same and verification methods are insufficient to compare the forecasts, we cannot directly compare the forecast performances among the RWCs. While comparing forecast performances among some RWCs is very informative, we have to proceed with efforts to compare the operational space weather forecasts. Verifying one's own forecast performance is the first step toward proceeding with the comparison of forecast performances. \par

The origin of the study of forecast verification goes back to 1884. Finley published some results of his tornado occurrence forecast (Finley 1884). Finley's results indicated that the accuracy of the tornado forecasts was extremely high, with the probability of correct forecasts exceeding 95\%. Shortly after the publication of Finley's paper, three papers that pointed out the deficiency of Finley's verification method appeared and alternative verification measures were proposed (Murphy 1996). These events were the start of forecast verification studies. Therefore, forecast verification has a long history in the terrestrial weather forecasting community, and the verification methods of the terrestrial weather forecast are more sophisticated than those of the space weather forecast. In this work, we perform a verification study of an operational solar flare forecast in RWC Japan (hereafter, RWCJ forecast) while referring to the methods of verifying terrestrial weather forecasts. \par

This article is organized as follows. In section \ref{sec:forecast}, we describe the RWCJ forecast and solar flare observation data. In section \ref{sec:verification}, we describe verification measures with the method of estimating the confidence intervals of the measures. The verification analyses of the RWCJ forecast are described in section \ref{sec:RWCJ}. A discussion and summary are given in sections \ref{sec:discussion} and \ref{sec:summary}, respectively. In the Appendix, brief descriptions of the definitions of the verification measures used in this study are shown. \par

\section{RWCJ forecast and flare observation data}
\label{sec:forecast}
The solar flare class is defined on the basis of the 1-8\AA\ X-ray flux observation by the Geostationary Operational Environmental Satellite (GOES). The RWCJ forecast gives the expected maximum solar flare class within 24 hours from the forecast issuing time. Table \ref{tbl:flare_class} shows the definition of the scale of the RWCJ forecast. We can easily recognize from the definition that the RWCJ forecast is a four-categorical deterministic forecast. Note that the flux level of a specific forecast has an upper bound, for example, when the forecast is ``Active'', the expected X-ray flux $F$ (W m$^{-2}$) is not $10^{-5}$ to infinity but $10^{-5}$ to $10^{-4}$. The RWCJ forecast is not an automatically determined forecast but a human decision forecast. Forecasters analyze many types of solar data such as the current situation and history of the solar X-ray flux, sunspot magnetic field configuration, and chromospheric brightening in active regions. Finally, the forecasters decide deterministically which class of solar activity will most likely occur in the next 24 hours. The forecast is issued at 6:00UT every day, so the range of the RWCJ forecast is from 6:00UT to 6:00UT the next day. While the RWCJ forecast started in 1992 and has continued to the present day, we use the RWCJ forecast data accumulated over 16 years, from 2000 to 2015 (5844 days), in this verification study. We did not use the data from 1992 to 1999 because a complete set of the forecast data could not be collected due to some missing data, and the forecast issue time was different until October 1994. \par

For solar flare observation data, we use solar activity event lists issued by SWPC/NOAA. The flare peak time and peak flux found in the event lists are used as the flare time and flare class, respectively. We define a day as from 6:00UT to 6:00UT the next day, which is the same as the RWCJ forecast range. The observed flare class of a day is defined as the maximum flare class of that day, the flare time of which is within the forecast range. Because of the automatic detection of X-ray flare for the event lists, some small and/or gradual increases in X-ray flux cannot be detected and listed as a flare even if it is a C-class activity. On 22 February 2001, for example, no flare was registered in the event lists although the flare activity was apparently C-class. Owing to the absence of flares, there were six days whose flare activities were incorrectly determined as below C-class activity instead of C-class. We corrected these incorrectly determined flare classes. By compiling the forecast and observation data, we obtained 5844 forecast-observation pair data that were analyzed to assess the RWCJ forecast performance. \par

\section{Forecast verification methods}
\label{sec:verification}
Here we introduce the verification methods for the RWCJ forecast. Using the 5844 forecast-observation pair data, we constructed a contingency table for the RWCJ forecast. Table \ref{tbl:ct} depicts a four-categorical contingency table for the RWCJ forecast. The number 0 to 3 on the forecast and observation axes stand for the flare class codes defined in Table \ref{tbl:flare_class}. \par

\subsection{Verification measures for dichotomous forecasts}
\label{sec:dichotomous_measure}
There are many scalar measures that can be used to verify the performance of a dichotomous forecast. Although the RWCJ forecast is not a dichotomous forecast, we can apply the RWCJ forecast to the scalar measures by collapsing the four-categorical forecast to a dichotomous forecast with a certain threshold. In this study, we deal with two types of event threshold. In the M-threshold, M-class or larger flares are defined as events and below M-class flares are defined as no events. In the X-threshold, X-class flares are defined as events and below X-class flares are defined as no events. \par

Many scalar verification measures have been proposed since the first paper on terrestrial weather forecast verification by Finley (1884). All scalar verification measures have some advantages and disadvantages, and it is not known which scalar verification measures are the most suitable for operational solar flare forecast verification although Bloomfield et al. (2012) recommended to use Peirce skill score for comparing a performance of solar flare forecasts. However, we may obtain some hints from terrestrial weather forecast verification strategies. The World Meteorological Organization (WMO) published a recommendation on verification methods for tropical cyclone forecasts, which describes the recommended scalar verification measures to assess tropical cyclone forecasts (WMO 2014). WMO (2009) published a recommendation for the verification and intercomparison of precipitation forecast, and this was reappeared in WMO (2014) together with an extremal dependence index (EDI), which has emerged as the importance for rare event forecast verification since WMO (2009). Because tropical cyclones are relatively rare events, tropical cyclone forecasts resemble large class solar flare forecasts in terms of their rarity. Table \ref{tbl:WMO} shows the recommended verification measures for assessing rare events that appeared in WMO (2014), who divided the scalar verification measures into three categories: mandatory, highly recommended, and recommended measures. The mandatory category is composed of hits, misses, false alarms, and correct rejections, which are all elements of a dichotomous forecast contingency table. The highly recommended measures are the frequency bias (FB), proportion correct (PC), probability of detection (POD), false alarm ratio (FAR), and equitable threat score (ETS). The recommended measures are the probability of false detection (POFD), critical success index (CSI), Peirce skill score (PSS), Heidke skill score (HSS), odds ratio (OR), odds ratio skill score (ORSS), and extremal dependence index (EDI). In this study, we estimate these verification measures except for the OR to assess the RWCJ forecast. Because the ORSS is the OR transformed so that its range is from $-1$ to $+1$, the OR and ORSS have a similar meaning. While two types of EDI were proposed, we use a symmetric version of the extremal dependence index (SEDI) because the SEDI has somewhat better properties than a non-symmetric version of EDI (Ferro \& Stephenson 2011). Brief definitions of these scalar verification measures are given in the Appendix 1. \par

We have to note that it is not sure whether this recommendation is the best for operational solar flare forecasting or not. An important implication of the recommendation is that only one verification measure is not enough to correctly assess the forecast performance (Murphy 1991), at least several attributes of forecast system such as bias, accuracy, discrimination, reliability, and skill, must be assessed by verification measures. \par

\subsection{Verification measure for multi-categorical forecast}
\label{sec:multi_measure}
As already mentioned, most of the scalar verification measures are defined on the basis of the dichotomous contingency table, so the multi-categorical contingency table must be collapsed to a dichotomous contingency table with a certain threshold so that the scalar verification measures can be applied to the RWCJ forecast. However, the collapsing process causes a loss of information included in the multi-categorical contingency table. Therefore, it is better to directly estimate a scalar verification measure for a multi-categorical forecast for the verification of the RWCJ forecasts. \par

Gandin \& Murphy (1992) proposed a scalar verification measure for multi-categorical forecasts. This measure is based on a scoring matrix, whose elements denote scores assigned to all elements of a multi-categorical contingency table, so it does not require the collapse of a multi-categorical contingency table. Moreover, the measure satisfies equitability, which is a highly desirable property for a scalar verification measure. An undesirable property of the measure is that it is not a closed form, meaning that some free parameters are required to estimate them. Gerrity (1992) derived a closed form of the Gandin \& Murphy-type verification measure, which does not require any free parameters. In this study, we use the scalar verification measure derived by Gerrity (1992), which we hereafter call the Gandin-Murphy-Gerrity score (GMGS). A brief definition of the GMGS is given in the Appendix 2. \par

A ranked probability score (RPS) is often used for a verification of a multi-categorical probabilistic forecast. If the forecast probabilities are set as one for any category and zero for others, it may be possible that the RPS is used for the verification of the multi-categorical deterministic forecast. However, as the RPS applied to a deterministic forecast does not satisfy equitability (see Appendix 3), the GMGS is suitable for a verification of a multi-categorical deterministic forecast. \par

\subsection{Confidence interval for verification measures}
\label{sec:ci}
Many of the forecast verification studies do not take data sampling uncertainty into account. However, because all of the verification measures are calculated from a finite number of sampled data, it is necessary to estimate confidence intervals for the calculated verification measures, as some authors have pointed out (e.g., Stephenson 2000; Jolliffe \& Stephenson 2003; Wilks 2006; Jolliffe 2007). \par

Because the elements of a dichotomous contingency table are regarded as binomial variables, the measures expressed as proportions such as the PC, POD, FAR, POFD, and CSI are sample estimates of the binomial probability $\hat p=x/n$. The simplest method of estimating a confidence interval of the binomial probability $\hat p$ is the so-called Wald confidence interval. The interval is calculated on the basis of a Gaussian approximation with mean $\hat p$ and variance $\sigma_p^2=\hat p(1-\hat p)/n$ instead of the binomial distribution. The resulting $1-\alpha$ confidence interval is 

\begin{equation}
	p=\hat p \pm z_{(1-\alpha/2)}\sqrt{\frac{\hat p(1-\hat p)}{n}},
\end{equation}
where $z_{(1-\alpha/2)}$ is the $1-\alpha/2$ quantile of the standard Gaussian distribution. The Wald confidence interval is simple, but can be rather inaccurate unless the number of sample $n$ is very large (Agresti \& Coull 1998). \par

A more accurate method of estimating a confidence interval was proposed by Wilson (1927). This confidence interval is also based on a Gaussian approximation, but its mean and variance are not assumed by the sample estimate of the binomial probability $\hat p$ but by using an unfixed value of the binomial probability as mean $p$ and variance $\sigma_p^2=p(1-p)/n$. This interval is derived from the inequality $p-z_{(1-\alpha/2)}\sigma_p\leq \hat p \leq p+z_{(1-\alpha/2)}\sigma_p$. The resulting $1-\alpha$ confidence interval is 

\begin{equation}
p=\frac{\hat p +\frac{z_{(1-\alpha/2)}^2}{2n}\pm z_{(1-\alpha/2)}\sqrt{\frac{\hat p(1-\hat p)}{n}+\frac{z_{(1-\alpha/2)}^2}{4n^2}}}{1+\frac{z_{(1-\alpha/2)}^2}{n}}.
\label{eq:score_ci}
\end{equation}
This confidence interval is sufficiently accurate even when the number of samples $n$ is quite small. According to Agresti \& Coull (1998), a confidence interval derived from later formula with sample number $n=5$ is more accurate than the Wald confidence interval derived with sample number $n=100$. For the confidence interval of the sample proportion, this formula is better than that for the Wald confidence interval. \par

There are some verification measures that cannot be written as sample proportions of the elements of the contingency table, such as the FB, ETS, HSS, PSS, ORSS, SEDI, and GMGS. For these verification measures, equation (\ref{eq:score_ci}) cannot be applied directly to estimate the confidence interval. In this case, an error propagation rule can be applied to the verification measures written as the function of the POD, POFD, and S (base rate), when estimating the confidence intervals. However, the error propagation rule assumes implicitly that confidence intervals of the POD, POFD, and S are sufficiently small with respect to a main values. In this verification study, although most of the verification measures have a sufficiently small confidence interval, some of the measures do not satisfy the assumption for a rare event forecast, in which case the error propagation rule cannot be applied to estimate the confidence interval. \par

To overcome these problems, we use a bootstrap method to estimate confidence intervals for scalar verification measures. The bootstrap method is becoming a popular means of constructing confidence intervals in statistical analyses and is based on a resampling procedure from an original data set. Many bootstrap replicates of our interest, such as scalar verification measures, are calculated from many resampled data sets, then their distribution is estimated from the many replicates. The confidence intervals are calculated by estimating a $1-\alpha/2$ quantile from the distribution. This means that an assumption about the distribution is not required in the bootstrap methods, which is a major advantage over the other methods. Details of the method have appeared in textbooks (e.g., Efron \& Tibshirani 1993). While there are many types of bootstrap confidence intervals, we use a BCa confidence interval in this study (BCa stands for ``bias-corrected and accelerated").  Because the BCa confidence interval includes corrections of bias and skew of bootstrap distribution, the BCa confidence interval has second-order accuracy\footnote{Other bootstrap confidence intervals with second-order accuracy, such as a bootstrap-{\it t} confidence interval and an ABC confidence interval, are also proposed (e.g., DiCiccio \& Efron 1996).} in the sample number $n$, whereas the simplest bootstrap confidence interval has an accuracy of the first order in $n$ (e.g., DiCiccio \& Efron 1996). Bootstrap samples, which are randomly sampled with replacement from original data set, are produced by Monte-Carlo method. A number of bootstrap samples is the same as the original data set (5844). A number of bootstrap replicates is 10,000 in this study. In statistical analyses, 95\% confidence intervals ($\alpha=0.05$) are often used, so we also estimate 95\% confidence intervals in this study. \par

\subsection{Distribution-oriented approach}
\label{sec:distribution-oriented}
In a distribution-oriented approach, a contingency table is regarded as a joint probability distribution for a pair of forecasts and observations, which is derived by dividing the elements of the contingency table by the total number of elements. The joint probability distribution is related to the attribute of the association between forecasts and observations. \par

A joint probability distribution can be factorized into a marginal distribution and a conditional probability distribution. There are two types of factorization in the forecast verification framework (Murphy \& Winkler 1987). One is a calibration-refinement factorization, and the other is a likelihood-base rate factorization. In the calibration-refinement factorization, the joint probability is factorized into the marginal probability of the forecasts and the conditional probability of the observations given the forecasts: $p(f,o)=p(o|f)\cdot p(f)$. In the likelihood-base rate factorization, the joint probability is factorized into the marginal probability of the observations and the conditional probability of the forecasts given the observations: $p(f,o)=p(f|o)\cdot p(o)$. The $p(o|f)$ and $p(f|o)$ are called the calibration distribution and likelihood distribution, respectively. The calibration distribution is related to attributes of reliability and resolution, whereas the likelihood distribution is related to the attribute of discrimination. Marginal distributions are related to the attribute of bias. Details of the distribution-oriented approach can be found in Murphy \& Winkler (1987). \par

\section{RWCJ forecast verification}
\label{sec:RWCJ}
\subsection{Verification of overall data}
\label{sec:overall}
In this section, we describe the results of the verification of the RWCJ forecast using data accumulated over 16 years. As mentioned in section \ref{sec:dichotomous_measure}, a multi-categorical forecast can be collapsed to a dichotomous forecast by setting a specific threshold when the conventional scalar verification measures are calculated. In this study, we deal with the M- and X-thresholds defined in section \ref{sec:dichotomous_measure}. Table \ref{tbl:verification_measure} summarizes the estimated scalar verification measures with a 95\% confidence interval for the M- and X-thresholds. \par

The scores of FB are almost unbiased in the M-threshold, meaning that the number of forecasted events is almost the same as the number of observed events. On the other hand, events are underforecasted in the X-threshold. \par

The accuracy of the RWCJ forecast seems to be extremely high and the X-threshold seems to be more accurate than the M-threshold according to the scores of PC. According to the numbers of hits and correct rejections in Table \ref{tbl:verification_measure}, most of the correct forecasts are forecasts of null events, and the null events can be forecasted easily when the event is quite rare. This is why the scores of PC are extremely high, especially for the X-threshold. On the other hand, the X-threshold is less accurate than the M-threshold in terms of the scores of CSI. As described in the Appendix 1, the correct rejections are not taken into consideration when calculating the CSI. This is why the scores of CSI shows the opposite trend to those of PC. When the correct forecasts of null events do not have an essential meaning, the CSI becomes a good measure of accuracy although an intuitive meaning of the measure is somewhat ambiguous. \par

The POD and POFD are the verification measures of discrimination, which is related to the likelihood distribution of the dichotomous contingency table. The discrimination means the ability of a forecast system to discriminate whether an event would occur under a condition of realized events. The scores of POD and POFD for the M- and X-thresholds are shown in Table \ref{tbl:verification_measure}. The POD for the M-threshold is reasonably good, meaning that the RWCJ forecast can discriminate the occurrence of below M-class flares and M- or X-class flares. On the other hand, the POD for the X-threshold is small, meaning that the RWCJ forecast cannot effectively discriminate the occurrence of below X-class and X-class flares. Because the POFD is a negative orientation measure, a small POFD means better performance. From the scores of POFD, it appears that the X-threshold has a better performance than the M-threshold. However, this is an invalid conclusion because the rarer the event, the larger the number of null events and the smaller the POFD. This means that the POFD approaches zero regardless of the discrimination performance when the event frequency approaches zero. \par

The reliability of a forecast can be expressed by the FAR. The FAR pertains to the relationship of a forecast to an average observation given the specific forecast and the calibration distribution of the dichotomous contingency table. The FAR is a negative orientation measure. For the RWCJ forecast, the X-threshold has bad performance, as shown in Table \ref{tbl:verification_measure}. 55-75 \% alarms issued for the occurrence of X-class flare were false alarms. This rate may be too large when the forecast is used for decision making to activate some countermeasures. \par

The ETS, HSS, and PSS are all verification measures of skill. All of these verification measures have significant positive values. All of these verification measures have larger scores for the M-threshold than for the X-threshold. It seems that the forecast skill of the M-threshold is better than that of the X-threshold. However, according to Stephenson et al. (2008), the scores of ETS, HSS, and PSS tend to degenerate to zero irrespective of their skill and behave as the trivial non-informative limit for vanishingly rare events. Therefore, we have to pay attention to the interpretation of the scores of the skill measures when comparing the scores for different event frequencies such as M- and X-thresholds. \par

The association describes the overall strength of the relationship between the individual pairs of forecasts and observations. According to the ORSS score, the association between the forecasts and the observations is reasonably strong for both the M- and X-thresholds. The ORSS for X-threshold is higher than that of M-threshold. This is easily accounted for from a definition of the ORSS (see Appendix 1). When the POFD is much smaller than the POD like the X-threshold, the ORSS approaches one. However, when the POFD is as same order of magnitude as the POD, the ORSS approaches zero even when POFD is small enough (see Figure \ref{fig:3modelX}). We note that the association and the accuracy seem to have the same attributes; however, they are different. The accuracy describes the correspondence between the individual pairs of forecasts and observations, while the association describes the overall strength of the relationship between them. \par

The SEDI was proposed by Ferro \& Stephenson (2011) and was designed to verify the performance of extremely rare event forecasts. The advantage of this measure is that the score for vanishingly rare events does not converge to zero but to a non-trivial meaningful value, whereas conventional measures such as the HSS, PSS, and ETS tend to be zero for vanishingly rare events. The scores of the SEDI in Table \ref{tbl:verification_measure} show that the RWCJ forecast for rare events is significantly better than random forecasting, for which the score of SEDI is zero. \par

\begin{figure}
	\centering
	\subfigure[Joint probability distribution]{
	\includegraphics[width=8cm]{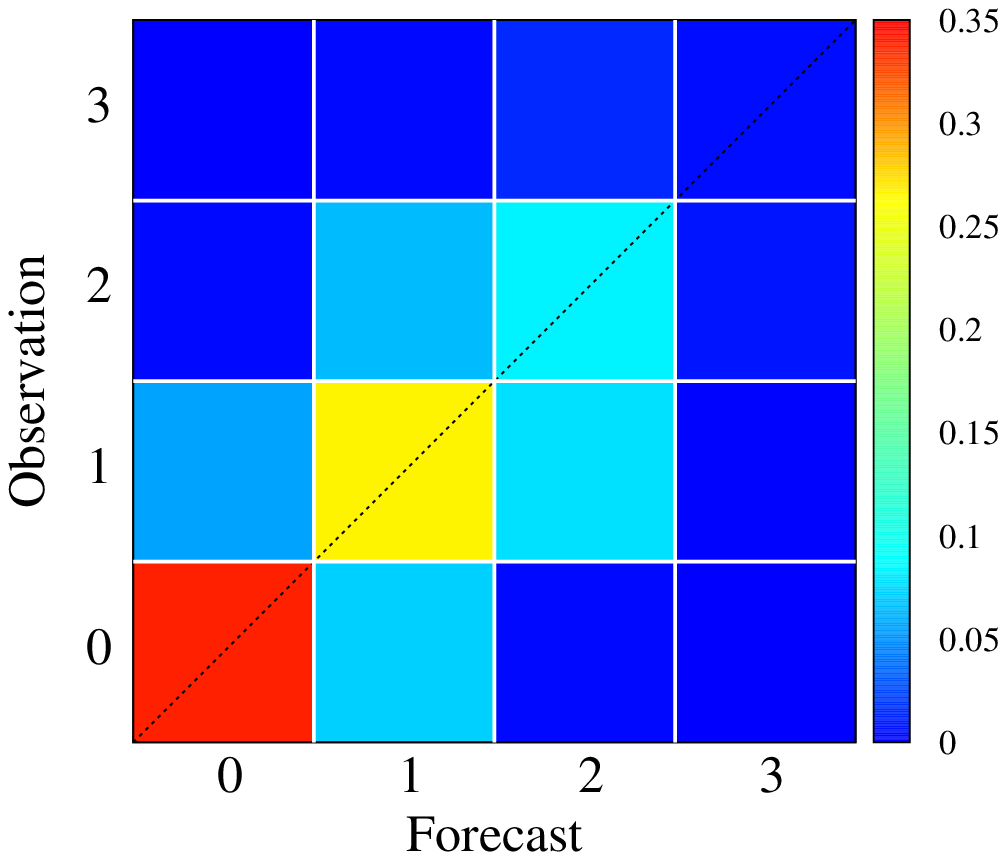}
	\label{fig:ct_joint}
	}
	\subfigure[Marginal distribution]{
	\includegraphics[width=8cm]{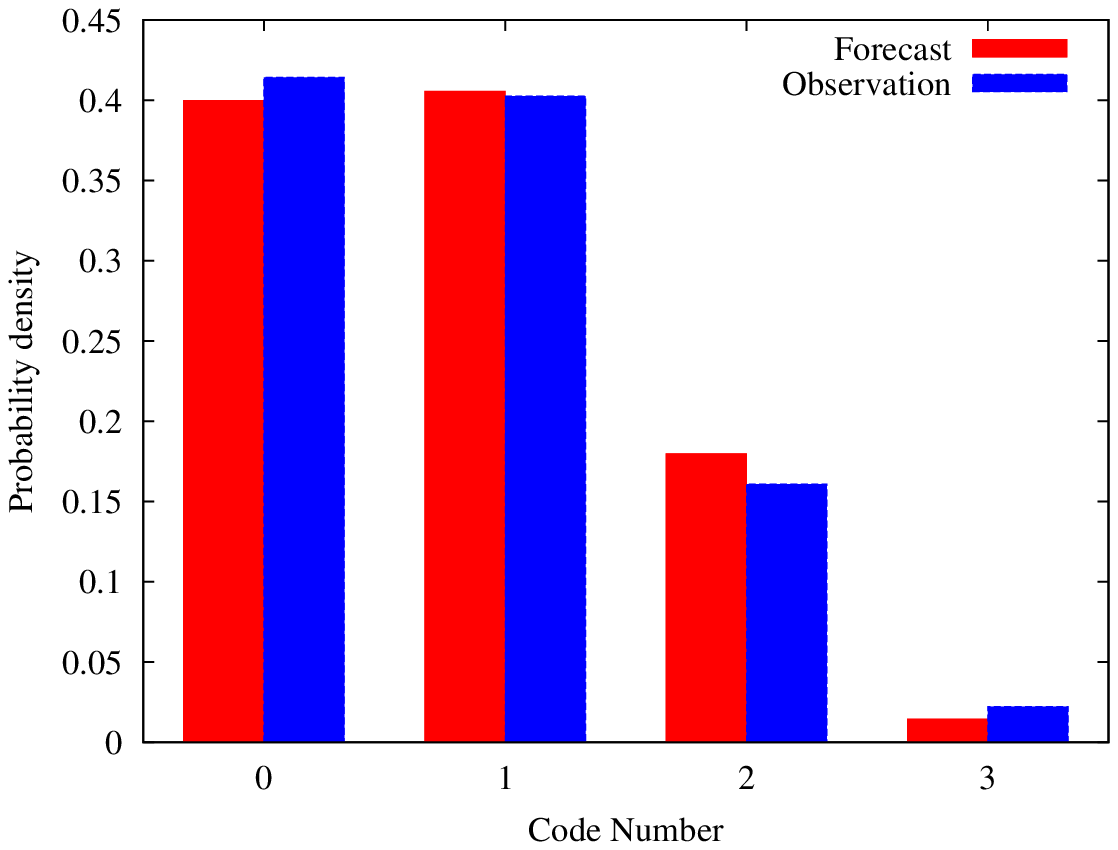}
	\label{fig:ct_marginal}
	}
	\subfigure[Calibration distribution]{
	\includegraphics[width=8cm]{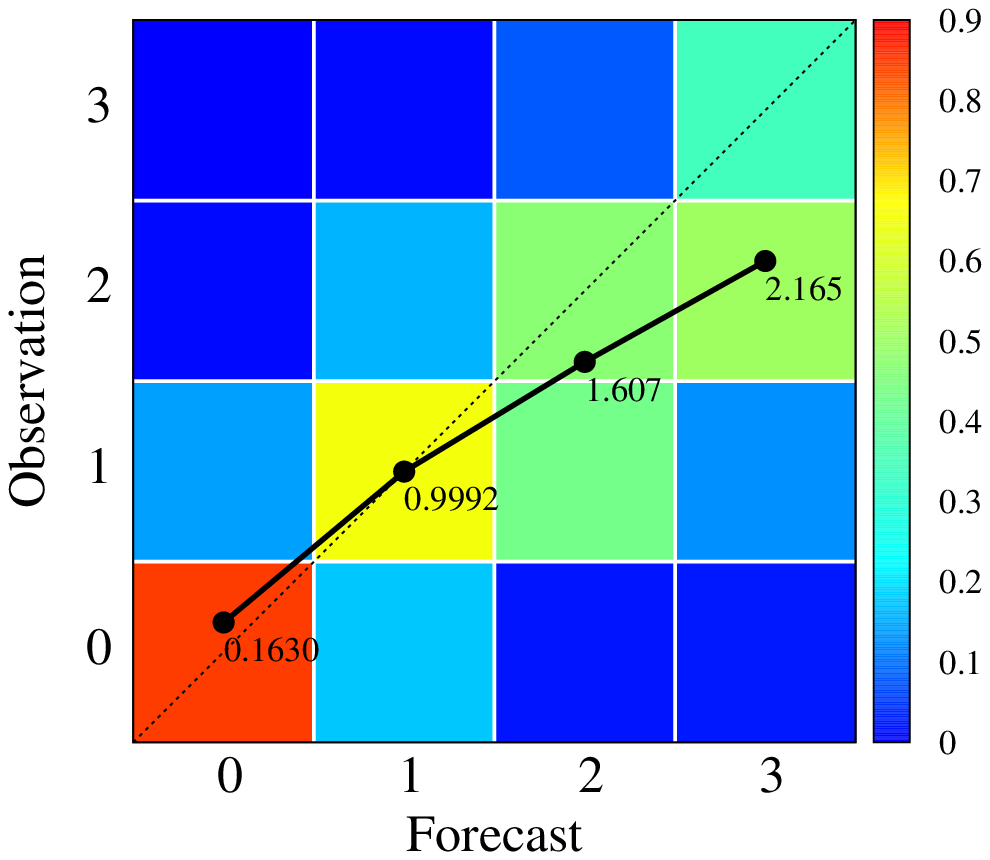}
	\label{fig:ct_calibration}
	}
	\subfigure[Likelihood distribution]{
	\includegraphics[width=8cm]{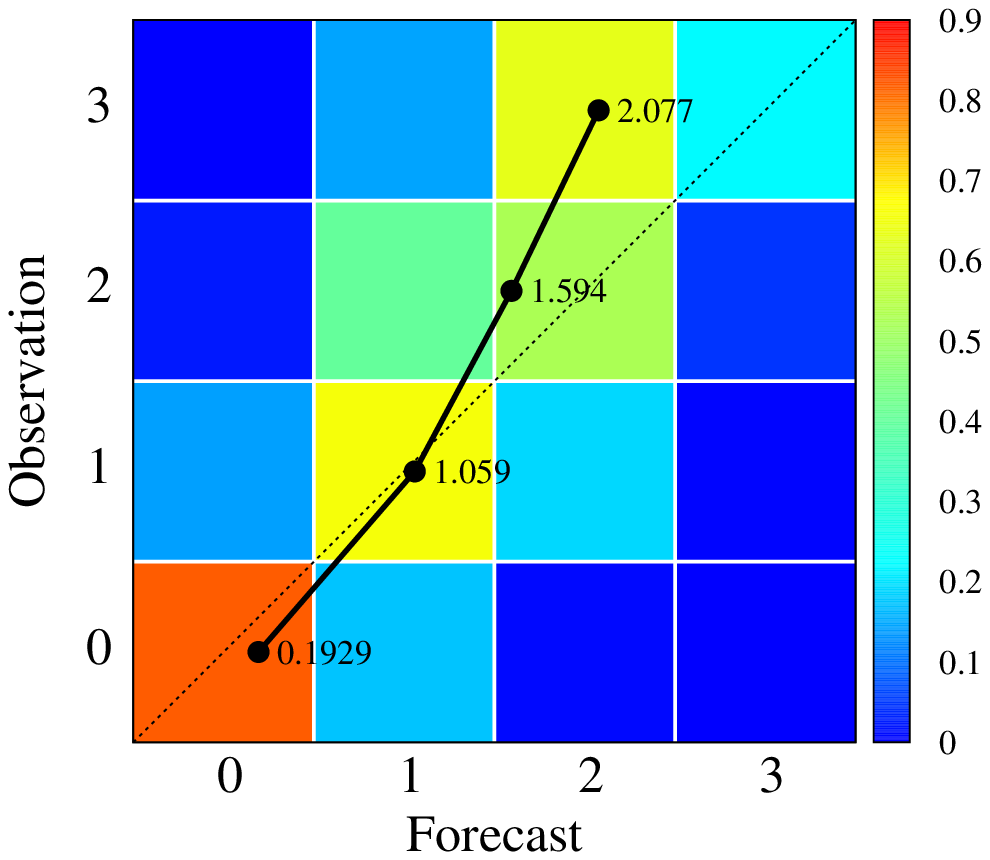}
	\label{fig:ct_likelihood}
	}
	\caption{Probability distributions for the RWCJ forecast. (a) Joint probability density. (b) Marginal distributions of forecast and observation. (c) and (d) Calibration and likelihood distributions, respectively. The black dots connected with lines in panels (c) and (d) stand for the conditional expectation values of the observation given the forecast and of the forecast given the observation, respectively.}
\end{figure}

In the rest of the subsection, a verification of the four-categorical forecast is given. Figure \ref{fig:ct_joint} depicts the joint probability distribution $p(f,o)$ for the RWCJ forecast calculated from the four-categorical contingency table in Table \ref{tbl:ct}. A good association between the RWCJ forecast and the observation can be seen in the figure. The correlation coefficient (CC) between them is estimated to be 0.717 with a 95\% confidence interval of [0.703, 0.730]. \par

Figures \ref{fig:ct_marginal} shows the marginal distributions of the RWCJ forecast and observation. The figure shows that the RWCJ forecast is almost unbiased, although there is slight overforecasting (underforecasting) for the M-class (X-class) flare. Figures \ref{fig:ct_calibration} and \ref{fig:ct_likelihood} show the calibration distribution and the likelihood distribution, respectively. The black dots with numbers connected by the line in Figure \ref{fig:ct_calibration} and \ref{fig:ct_likelihood} are the conditional expectation values of the observation given the forecast and of the forecast given the observation, respectively. We can recognize from Figure \ref{fig:ct_calibration} that the flare class that most frequently occurred under the condition that a specific flare class had been forecasted was the same as the forecasted flare class except when an X-class flare had been forecasted. When X-class flares had been forecasted, the most frequently occurring flares were M-class flares. This means that the reliability of the RWCJ forecast is good for below X-class flares but not good for X-class flares. As shown in Figure \ref{fig:ct_likelihood}, under the condition that a specific class flare occurred, the most frequently forecasted flare class was the same as the observed class except for the condition that an X-class flare occurred. The most frequently forecasted flare class under the condition that an X-class flare occurred was an M-class flare. This means that the RWCJ forecast cannot successfully discriminate between the occurrences of X-class and below X-class flares. \par

The GMGS was calculated from the four-categorical contingency table in Table \ref{tbl:ct}. As already mentioned, because the GMGS satisfies equitability, the score for unskillful forecasts becomes zero. For the RWCJ forecast shown in Table \ref{tbl:verification_measure}, the score is significantly larger than zero, meaning that the RWCJ forecast has some forecast skill. \par

\subsection{Comparison with persistence and recurrence method}
\label{sec:comp}
It was shown in section \ref{sec:overall} that the RWCJ forecast seems to have reasonable performance. In this subsection, we compare the performance of the RWCJ forecast with those of two other forecasting methods: a persistence method and a recurrence method. In the persistence method, today's forecast is the same as yesterday's observation results. In the recurrence method, today's forecast is the same as the observation results 27 days ago. As solar flare activity is not independent between consecutive days, it is expected that the persistence method have some forecast performance. As the solar rotation period when viewed from Earth is almost 27 days, the recurrence method is also expected to have some forecast performance. Therefore, it is useful to compare the three forecasting methods to assess the RWCJ forecast performance. \par

\begin{figure}
	\includegraphics[width=12cm]{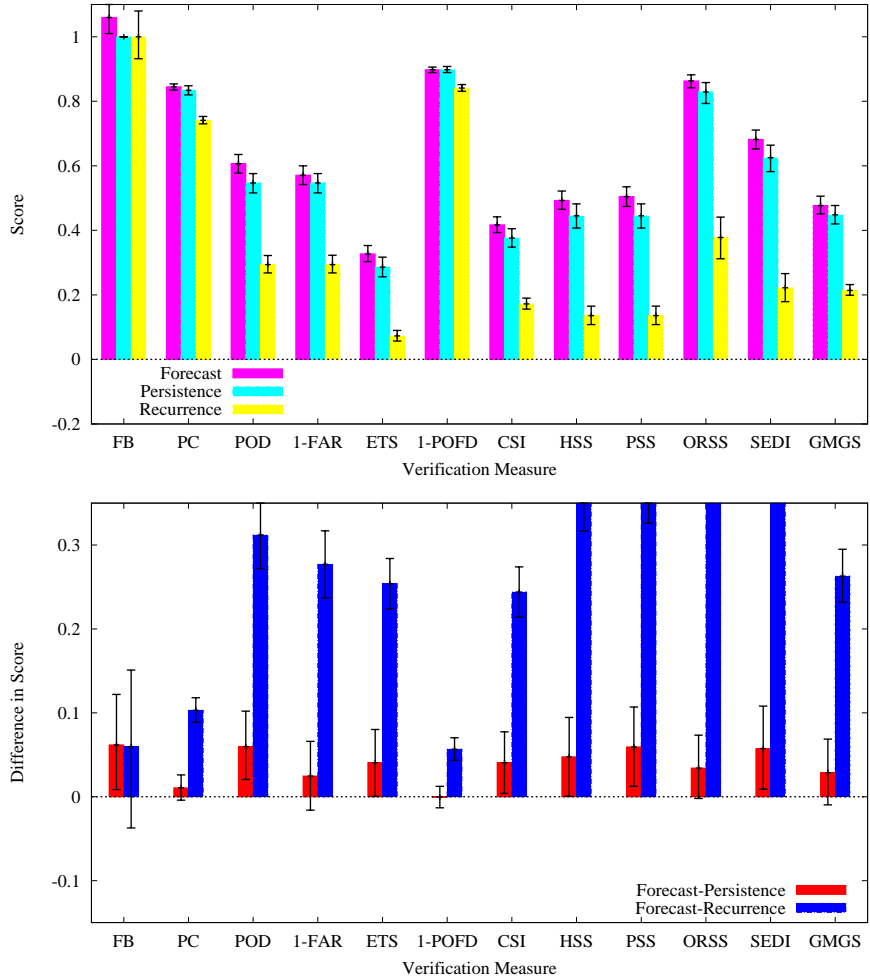}
	\caption{Comparison of various verification measures among the RWCJ forecast, persistence method, and recurrence method for the M-threshold. The top panel shows absolute scores of the verification measures. Magenta, cyan, and yellow stand for the RWCJ forecast, persistence method, and recurrence method, respectively. The bottom panel shows the difference in the scores between the RWCJ forecast and the persistence method as red and that between the RWCJ forecast and the recurrence method as blue. The black bars stand for the 95\% confidence intervals of the scores. We note that as the FAR and POFD are negative orientation measures, the 1-FAR and 1-POFD are plotted, which are positive orientation values.}
	\label{fig:3modelM}
\end{figure}

The top panel of Figure \ref{fig:3modelM} shows comparisons of various verification measures among the three forecasting methods for the M-threshold. The magenta, cyan, and yellow bars show the resultant scores for RWCJ, persistence, and recurrence methods, respectively. The black intervals drawn on the colored bars stand for the 95\% confidence intervals of the scores. We can easily recognize that all three methods are skillful forecasts because all the scores of ETS, HSS, and PSS are positive. All the verification measures, except for FB, for the recurrence method have significantly worse scores than the RWCJ forecast (FAR and POFD are negative orientation measures), meaning that the performance of the RWCJ forecast is significantly better than that of the recurrence method. On the other hand, the differences in the scores between the RWCJ forecast and the persistence method are small for all the verification measures, and the 95\% confidence intervals overlap each other. To investigate the difference in scores between the RWCJ forecast and persistence method more precisely, we show the difference in the various scores between the RWCJ forecast and the other two methods with 95\% confidence intervals in the bottom panel of Figure \ref{fig:3modelM}. The red bars stand for the differences in scores between the RWCJ forecast and the persistence method while the blue bars stand for the differences between the RWCJ forecast and the recurrence method. We can recognize that all the verification measures have slightly better scores for the RWCJ forecast than for the persistence method. However, the 95\% confidence intervals for some measures such as the PC, FAR, POFD, and ORSS include the zero score. This means that we cannot definitely conclude that there are significant differences in the scores between the RWCJ forecast and the persistence method. What we can conclude from the results for the performances of the RWCJ forecast and the persistence method is that (1) there is no significant difference in the accuracy, but the accuracy is slightly better when correct forecasts of a null event are not essential, (2) discrimination is slightly better for the RWCJ forecast, (3) there is no significant difference in reliability, (4) the RWCJ forecast has slightly better skill than the persistence method, (5) there is very slight or no significant difference in association, and (6) the performance of extreme event forecasts is slightly better for the RWCJ forecast. In summary, the RWCJ forecast for the M-threshold seems to have a slightly better performance than the persistence method. \par

\begin{figure}
	\includegraphics[width=12cm]{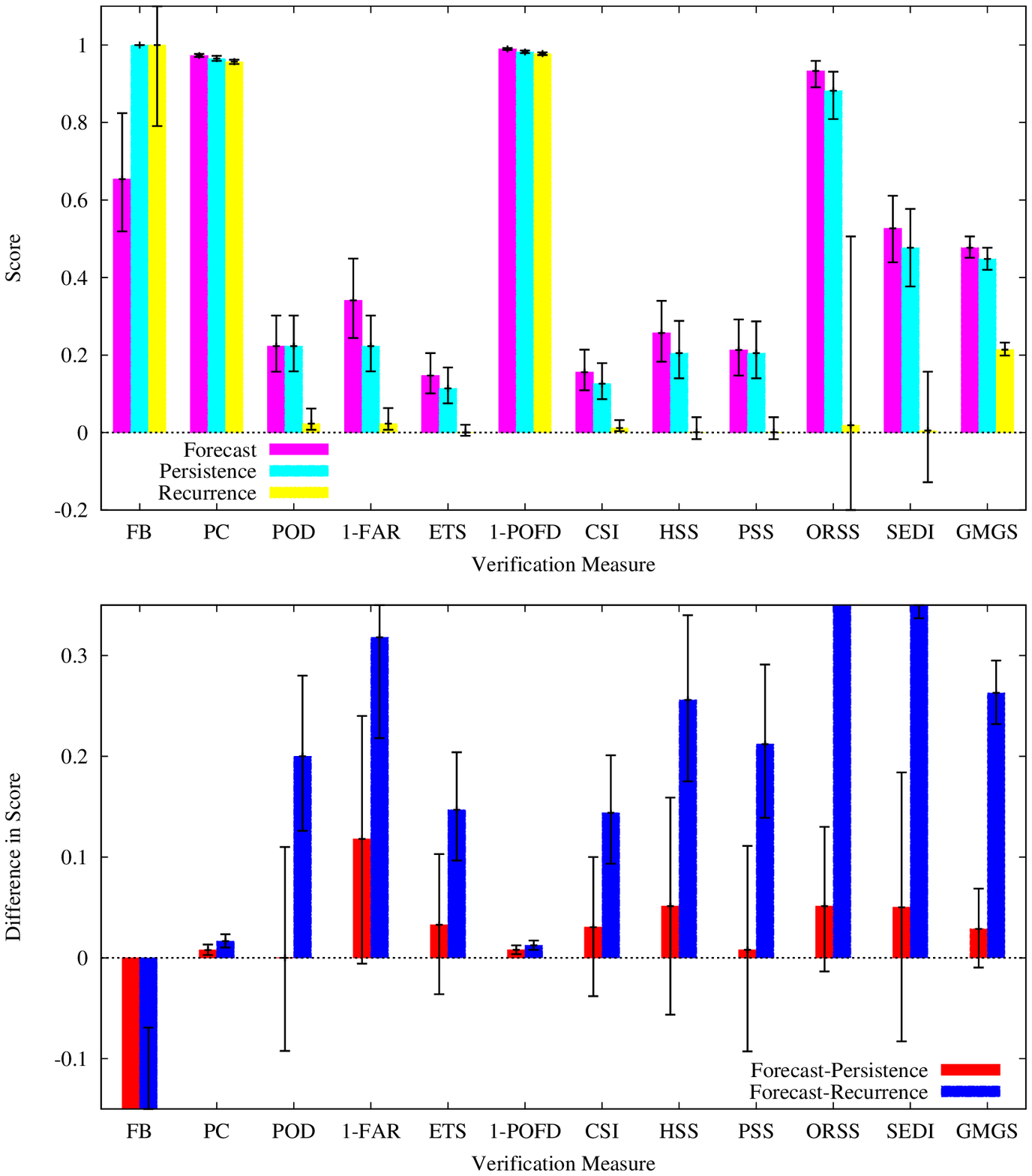}
	\caption{Same as Figure \ref{fig:3modelM}, except that the event definition is the X-threshold.}
	\label{fig:3modelX}
\end{figure}

Figure \ref{fig:3modelX} is the same as Figure \ref{fig:3modelM} except that the event definition is the X-threshold. From the top panel of Figure \ref{fig:3modelX}, we can immediately recognize that the RWCJ forecast and persistence method have some forecast performance, whereas the recurrence method has no performance. This may mean that the recurrence of an active region that produced an X-class flare during the last solar rotation period provides little information on the productivity of the X-class flare. Differences in scores between the RWCJ forecast and the persistence method are shown as red bars in the bottom panel of Figure \ref{fig:3modelX}. For almost all verification measures, the 95\% confidence intervals include the zero score, meaning that we cannot definitely conclude that there is a significant difference between the performances of the RWCJ forecast and the persistence method for the X-threshold. \par

The scores of GMGS for the three forecast methods are included in Figures \ref{fig:3modelM} and \ref{fig:3modelX}. The scores drawn in these two figures are exactly the same because the GMGS is calculated directly from the four-categorical contingency table, not a collapsed dichotomous contingency table. Therefore, the measures express the performances of the four-categorical forecast system. The scores of the GMGS show that all three forecast methods have some skill as a four-categorical forecast system. The skill of the recurrence method is significantly worse than that of the other two methods. The difference between the RWCJ forecast and the persistence method may not be significant. \par

\subsection{Verification of subset data}
\label{sec:subset}
\begin{figure}
	\includegraphics[width=12cm]{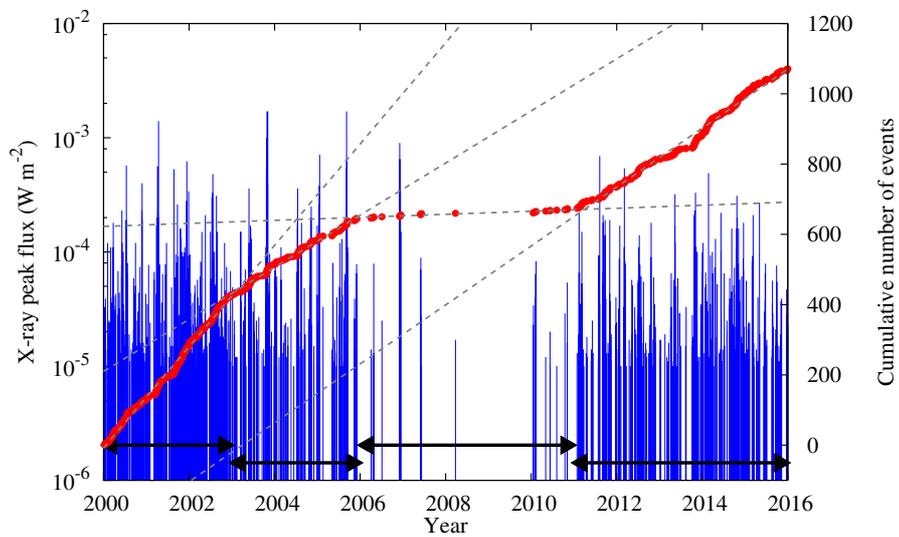}
	\caption{Chronological history of the M-threshold event occurrence. The blue vertical lines show the maximum M-threshold events within 24 hours from the forecast issue time. The red dots show the cumulative number of maximum M-threshold events plotted against the event occurrence date. The gray dashed lines are fitted lines of cumulative number of events for each subsets.}
	\label{fig:event_freq}
\end{figure}

Hamill \& Juras (2006) revealed that some conventional scalar verification measures such as the ETS were prone to give an unexpectedly increased score when the climatological frequency of the event occurrence varies among pooled samples. In the case of solar flare forecasts, the climatological frequency of event occurrence appears to vary chronologically owing to changing solar activity. Therefore, we apply the verification study to subset data, which are divided by solar activity levels. Figure \ref{fig:event_freq} depicts a chronological history of the M-threshold event occurrence defined in section \ref{sec:overall}. The blue vertical lines show the maximum M-threshold solar flare events within 24 hours from the forecast issue time. The red dots show the cumulative number of maximum M-threshold events counted from 1 January 2000 plotted against the event occurrence date. The slope of the red dots can be regarded as the climatological frequency of event occurrence. We can clearly recognize that there are four separate periods, during which the event frequencies are almost constant, which are shown by the dashed gray lines. Therefore, we divide the 16 years of data into four subsets: 2000-2002 (subset-1), 2003-2005 (subset-2), 2006-2010 (subset-3), and 2011-2015 (subset-4). \par

\begin{figure}
	\includegraphics[width=12cm]{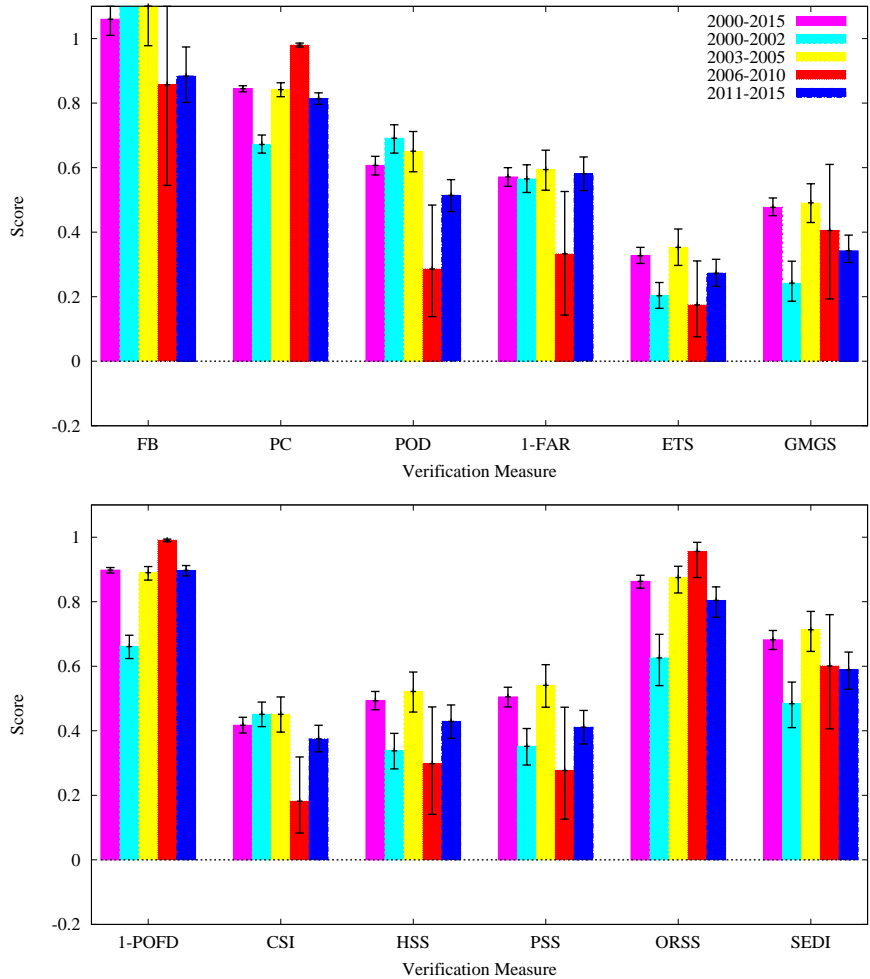}
	\caption{Comparison of various verification measures with 95\% confidence interval for the M-threshold for four subsets with the overall data as a reference. The magenta, cyan, yellow, red, and blue bars stand for 2000-2015 (overall), 2000-2002 (subset-1), 2003-2005 (subset-2), 2006-2010 (subset-3), and 2011-2015 (subset-4), respectively. We note that as the FAR and POFD are negative orientation measures, the 1-FAR and 1-POFD are plotted, which are positive orientation values.}
	\label{fig:subsetM}
\end{figure}

\begin{figure}
	\includegraphics[width=12cm]{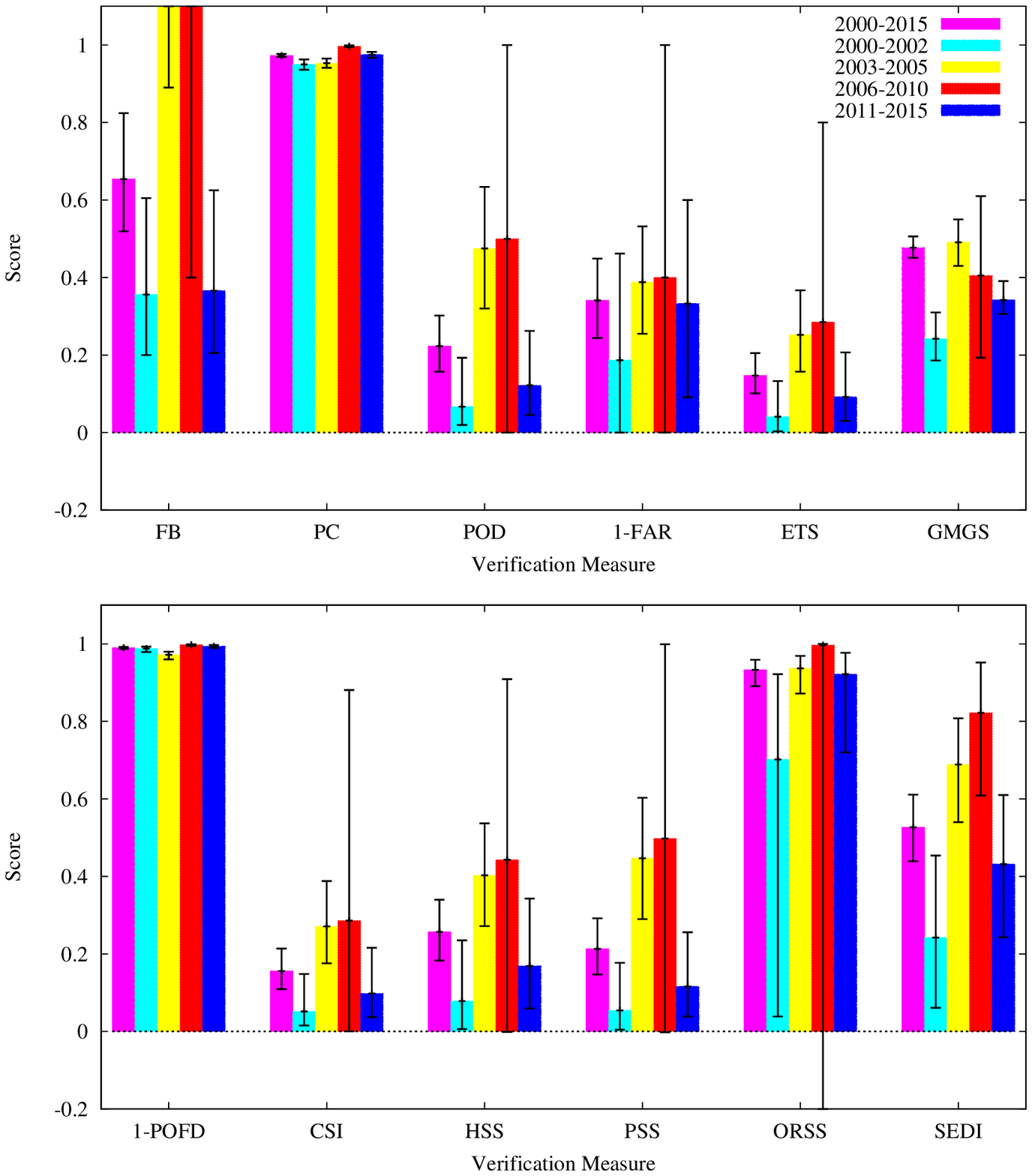}
	\caption{Same as Figure \ref{fig:subsetM}, except that the event definition is the X-threshold.}
	\label{fig:subsetX}
\end{figure}

Figures \ref{fig:subsetM} and \ref{fig:subsetX} show the scores of various verification measures for the four subsets with a whole dataset as references for the M- and X-thresholds, respectively. The cyan, yellow, red, and blue bars are for subset-1 through subset-4, respectively, with magenta used for the whole dataset. In the M-threshold, the FB shows that subset-1 and 2 are overforecasting while subset-3 and 4 are underforecasting, although subset-3 has very wide 95\% confidence interval. According to the PC, accuracy seems to be the best for subset-3. However, the high score of subset-3 is ascribed by correct forecast of null events because the CSI of subset-3 is significantly lower than that of other three subsets. Comparing the subset-2 and 4, which have almost the same event frequency, discrimination shown by the POD of subset-2 is better than that of subset-4. From the POD and FAR, we can recognize that subset-3 has low discrimination and reliability than other three subsets. For the forecast skill, we can see from the ETS, HSS, and PSS that subset-2 seems to have the best forecast skill among the four subsets, although a slight overlap of the 95\% confidence intervals exists. The SEDI shows that the performance of extreme event forecast is subset-2, 4, and 1 in descending order, except for subset-3 which have large uncertainty. In the X-threshold, we can recognize from the FB that subset-1 and 4 are large underforecasting. Accuracy shown by the CSI is subset-2, 4, and 1 in descending order, although 95\% confidence intervals overwrap each other. Low PODs for subset-1 and 4 would be ascribed by large underforecasting. There seems to be no significant difference for reliability among all subsets because all the 95 \% confidence intervals of FAR overwrap each other. It seems that subset-2 has the best forecast skill among the subsets except for subset-3, for which we cannot comment on the forecast skill because the 95\% confidence intervals for subset-3 are extremely wide. The SEDI has the same pattern as the POD because the POFDs for all subsets have almost the same values and the SEDI is calculated by only the POD and POFD. Regarding the skill of four-categorical forecast, the GMGS shows that the best forecast skill seems to be for subset-2, although the upper limit of the relatively wide confidence interval of subset-3 is higher than that of subset-2. \par

\begin{figure}
	\includegraphics[width=12cm]{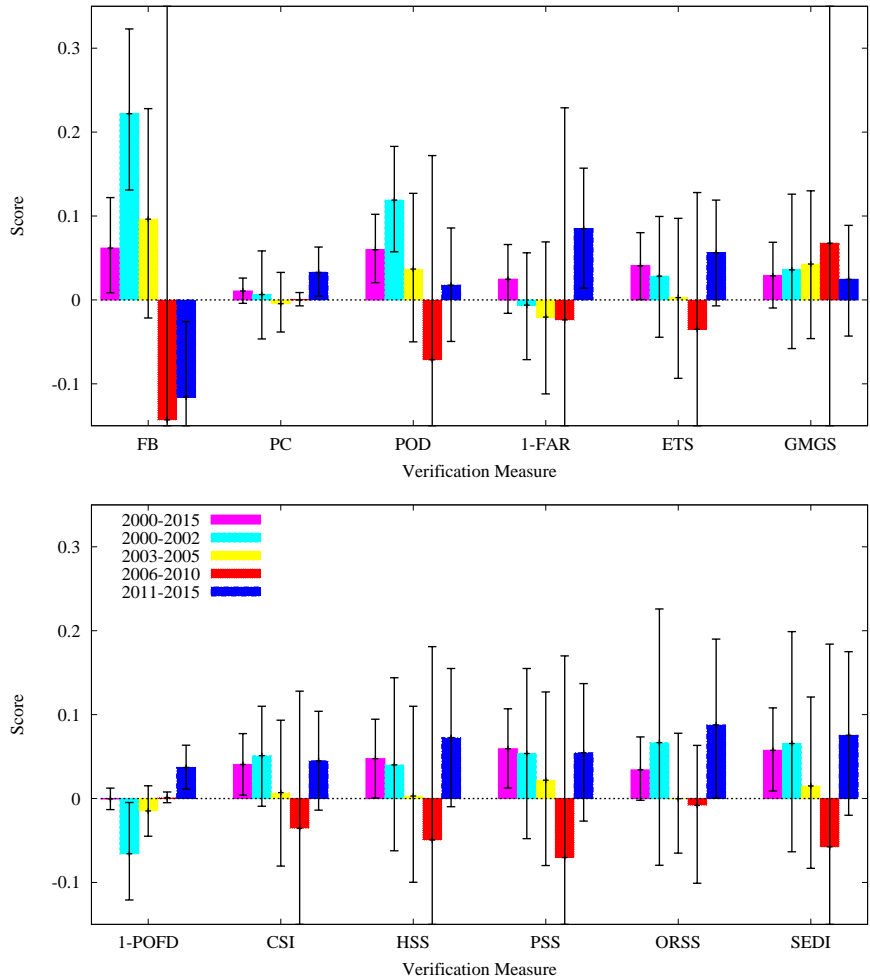}
	\caption{Difference in scores with 95\% confidence interval for M-threshold event between the RWCJ forecast and the persistence method. The magenta, cyan, yellow, red, and blue bars stand for 2000-2015 (overall), 2000-2002 (subset-1), 2003-2005 (subset-2), 2006-2010 (subset-3), and 2011-2015 (subset-4), respectively. We note that as the FAR and POFD are negative orientation measures, the 1-FAR and 1-POFD are plotted, which are positive orientation values.}
	\label{fig:subset_diffM}
\end{figure}

As already shown in section \ref{sec:comp}, the skill score of the RWCJ forecast is often similar to that of the persistence method. This may imply that a relatively high score of subset-2 is due to the persistence of event occurrence because the solar activity during the period of subset-2 was reasonably high. To investigate this point, the differences in score between the RWCJ forecast and the persistence method for each subset for the M-threshold are drawn in Figure \ref{fig:subset_diffM}. For most of the verification measures, a zero score is included in the 95\% confidence interval, meaning that we cannot conclude that there are significant differences between the RWCJ forecast and the persistence method for all the subsets. The comparison of each subset for a specific verification measure is also important. For the verification measure of the skill (ETS, HSS, and PSS), the largest differences in scores appear in subset-4. For subset-2, the difference in score is positive but the smallest among all the subsets except for subset-3. Although we cannot give a definitive conclusion because the difference among the subsets is small and the confidence intervals overlap each other, the relatively high score of subset-2 is probably due to the persistence of event occurrence. Moreover, the best {\it judgment} skill, which is defined in Section \ref{sec:discussion}, for the M-threshold may be in subset-4 because the difference in scores of the ETS, HSS, and PSS between the RWCJ forecast and the persistence method for the subset-4 is the largest among those for all subsets. In the X-threshold, we cannot comment on the significance of the score differences because of the extremely wide confidence intervals compared with the score difference (not shown). \par

\section{Discussion}
\label{sec:discussion}
As already mentioned in section \ref{sec:comp}, the persistence method, as well as the RWCJ forecast, is a skillful forecast method. However, the persistence method is determined by only the observation result of the previous day, namely, the persistence method is a skillful forecast method {\it without} judgment. On the other hand, the RWCJ forecast (and other operational solar flare forecasts) includes a judgment process to determine the issued forecast. If the skill scores of the operational solar flare forecast are smaller than those of the persistence method, the contribution of the judgment of flare occurrence to the operational solar flare forecast is almost zero. Therefore, it would be better to assess the {\it judgment} skill for the operational solar flare forecast in addition to forecast skills, which are assessed by the verification measures of ETS, HSS, and PSS. In sections \ref{sec:comp} and \ref{sec:subset}, we estimated the difference in the scores between the RWCJ forecast and the persistence method as one of the assessments of the judgment skill. It is also useful for estimating a skill measure with reference to the persistence method to assess the judgment skill (called a judgment skill measure henceforth). Similar to the HSS, the judgment skill measure (JS) based on the PC is defined as 

\begin{equation}
	JS=\frac{a-a_p+d-d_p}{a-a_p+b+c+d-d_p}=\frac{PC-PC_p}{1-PC_p},
\end{equation}
where $a$, $b$, $c$, and $d$ are the elements of the dichotomous contingency table (see, Table \ref{tbl:dichoto_table}). The elements with subscript ``$p$'' are for the persistence method. The scores of the JS are 1 and 0 for a perfect forecast and the persistence method, respectively. However, for a perfectly incorrect forecast the score of the JS is not -1 but depends on the score of the persistence method. One of the most important characteristics that a verification measure of skill should have is equitability (Gandin \& Murphy 1992). The equitability means that the unskillful forecasts involving forecasting ``yes'' every time, forecasting ``no'' every time, and forecasting at random must have the same scores. However, the JS does not satisfy equitability. This is not a good characteristics for a verification measure of skill. Recognizing this characteristic, it is better to use the judgment skill measure JS. For the RWCJ forecast for overall data, the JS is estimated to be 0.0639 [-0.0305, 0.145] for the M-threshold and 0.223 [0.0843, 0.332] for the X-threshold. Since the 95\% confidence intervals of the scores of JS for M- and X-threshold are largely overwrapped, the difference in the JS scores are probably not significant. We have to pay attention to interpretation of the JS for different event thresholds in case of PC$\sim$PC$_p$, because a denominator of the JS is dominated by the number of false alarms ($b$) and misses ($c$), which can be small for rare event forecast, and the JS can be large irrespective of their judgment skill. Similar to the ETS, the judgment skill measure based on the CSI can also be considered, whose definition is the same as the ETS except with $a_r$ replacing $a_p$. However, this formulation has a critical defect. When the judgment skill is very poor, it may be possible that $a_p$ is larger than $a+b+c$. In this case, the defined formula is positive despite the very poor judgment skill. Therefore, the formulation based on the CSI cannot be used to assess the judgment skill. \par

As the RWCJ forecast is a four-categorical forecast, the performance assessment of only the collapsed dichotomous forecast is insufficient, and the direct assessment of a four-categorical contingency table is required in addition to the verification as a dichotomous forecast with some thresholds. However, a small scalar verification measure for a multi-categorical contingency table has been discovered, for example, the GMGS for measure of skill. For accuracy, we can use a proportion correct extended to a multi-categorical contingency table (PC$_{\mathrm m}$), which is defined as PC$_{\mathrm m}$=$\sum_i p_{ii}$, where $p_{ii}$ is the diagonal element of a joint probability distribution for the multi-categorical contingency table (e.g., Jolliffe \& Stephenson 2003). No scalar verification measures for other attributes such as reliability, resolution, and discrimination have been proposed yet. Therefore, a distribution-oriented approach such as the discussion of a joint probability distribution in section \ref{sec:overall} is also required. We propose a verification strategy for a multi-categorical deterministic operational solar flare forecast as follows: marginal distributions for bias, PC$_{\mathrm m}$ for accuracy, CC and joint probability distribution for association, the likelihood distribution for discrimination, the calibration distribution for reliability and resolution, the GMGS for forecast skill, and JS$_{\mathrm m}$ for judgment skill, whose definition is the same as the JS except with PC replaced by PC$_{\mathrm m}$ and PC$_{\mathrm p}$ replaced by PC$_{\mathrm{mp}}$, in addition to the verification as dichotomous forecasts with M- and X-thresholds. The scores of the proposed scalar verification measures for the overall data of multi-categorical RWCJ forecast are summarized in Table \ref{tbl:multi_score}. The accuracy as the four-categorical forecast is reasonably good because PC$_{\mathrm m}$ is 68\% to 71\%. According to CC and joint probability distribution, the association for four-categorical forecast is also reasonably good. However, forecast skill is not so high although it may be acceptable depending on the user needs. As the GMGS imposes larger penalty for multi-category error than for one-category error, reducing the multi-category error will lead high score. Although the confidence interval of the JS$_{\mathrm m}$ does not include zero, the judgment skill as four-categorical forecast is small. \par

On the other hand, the dichotomous forecast verification strategy for a deterministic operational solar flare forecast is under discussion. As already mentioned, many verification measures for a dichotomous deterministic forecast have been proposed since the paper by Finley (1884). Because all verification measures have both advantage and disadvantage, it is difficult to determine which verification measure is the best for operational solar flare forecasting. As we already mentioned, the CSI will be more suitable than the PC for rare event forecasting. However, an intuitive meaning of the CSI is somewhat ambiguous, while a meaning of the PC is completely clear, which means a percentage of correct forecast over whole forecast. For a verification measure of skill, the ETS is commonly used in terrestrial weather forecasting community, while the PSS is recommended by Bloomfield et al. (2012) in the space weather forecasting community. The PSS is base-rate-independent measure, which is a good characteristic for verification measure of skill, while the ETS is base-rate-dependent measure. On the other hand, while the PSS are designed on the basis of the PC, the ETS is defined on the basis of the CSI. So, it is a difficult question which measure is more suitable for verifying rare event forecasts. These kinds of discussions appear frequently for other verification measures. With the understanding of this situation, we propose a verification strategy for dichotomous forecast for rare event as follow: FB for bias, PC and CSI for accuracy, POD for discrimination, FAR for reliability, PSS for forecast skill, and SEDI for association. As the SEDI is formally a non-linear transformation of the OR (Ferro \& Stephenson 2011) although its derivation is completely different, the SEDI can be regarded as the verification measure of association for rare event forecast. For the verification measure of skill, although there are many discussions for their usability, the PSS is selected for respecting Bloomfield et al. (2012). We have to note that this is just one suggestion. We think further researches on verification measures themselves are highly required to determine the best verification strategy for dichotomous deterministic operational solar flare forecast.

As already mentioned in Section \ref{sec:intro}, comparing forecast performances among some RWCs is informative. We briefly discuss comparison between the RWCJ forecast and the RWC Belgium (ROB) solar flare forecast (Devos et al., 2014). As a verification study of RWC USA (Crown 2012) is for probabilistic forecast of solar flare, it cannot be compared directly with RWCJ forecast. Scores for some verification measures for M-threshold for ROB are shown in Table 3 of Devos et al., (2014). Comparison between scores for ROB and RWCJ forecast (Table \ref{tbl:verification_measure} in this article) shows following results: (1) RWCJ is almost unbiased while ROB is obviously under-forecasted, (2) accuracy of RWCJ is a little worse than that of ROB according to PC, however, excluding trivial correct forecasts (correct rejections) leads reverse result (CSI of ROB is estimated as 0.311), (3) discrimination of RWCJ is obviously better than that of ROB according to POD, (4) reliability of RWCJ is somewhat worse than that of ROB because FAR of RWCJ is larger than that of ROB, (5) forecast skill of RWCJ is somewhat better than that of ROB, (6) performance of extreme event forecast of RWCJ is slightly better than that of ROB because SEDI of ROB is estimated as 0.594 from their Table 3. The obvious under-forecasting of ROB may lead the small POD and FAR. Because a large over-forecasting (under-forecasting) leads sometimes a large POD and FAR (small POD and FAR), an unbiased forecast will be required for a good forecasting system. We should stress that a period of verified data used in Devos et al., (2014) is different from that of this study. \par

RWCJ forecast is a four-categorial deterministic forecast which is defined by ISES. However, some readers may think that only M- and X-threshold forecasts are enough because B- or C-class flares will not affect social infrastructures. This claim may be true. On the other hand, the M- and X-threshold forecasts can be easily made by setting some threshold in the four-categorical forecast. Of course, no (or almost no) flare forecast such as less than or equal to B-class flare forecast can be made from the four-categorical forecast by setting the threshold between categories of B- and C-class flares. Therefore, various threshold forecasts, which the forecast users require, can be made from the four-categorical forecast. As many RWCs belonging to ISES have issued four-categorical forecast, issuing the four-categorical forecast is preferable for the forecast verification researches among some RWCs. Another preferable forecast is a probabilistic forecast. Because deterministic solar flare forecast based on the underlying physics is unlikely to be realized, a probabilistic forecast is essential (Kubo 2008). We think that a future direction of our operational solar flare forecast will be a probabilistic forecast. \par

Forecast performance comparison with a persistence and recurrence method was performed in this study. This is a common approach as a verification study of space weather as well as terrestrial weather forecasting, for example, Devos et al. (2014). This approach is also justified by the scientific researches of solar flare occurrence. McCloskey et al. (2016) showed that time evolution of flare occurrence is important factor for flare forecasting. This fact will be related with the persistence method to be skillful. D\'emoulin et al. (2002) and Green et al. (2002) showed by analyzing long lifetime active regions that magnetic flux and helicity stored in an active region, which are expelled by solar flares accompanied by coronal mass ejections, are often reduced during successive solar rotations. This fact may be related with that the recurrence method is almost unskillful forecast, which was shown by verification of X-threshold for recurrence method in this study. \par

\section{Summary}
\label{sec:summary}
A verification study on the operational solar flare forecast in the Regional Warning Center Japan (RWCJ forecast) was performed for the first time. Forecast and observation pair data accumulated over 16 years from 2000 to 2015 were used in the study. We estimated various types of scalar verification measures with 95\% confidence intervals for the overall data of the RWCJ forecast, and they were compared with those of persistence and recurrence methods. The performance of the recurrence method is significantly worse than that of the RWCJ forecast. However, the score difference in various verification measures between the RWCJ forecast and the persistence method is small, and we could not conclude definitely that there were significant performance differences between these two forecast methods, although a slightly significant difference was found for the M-threshold events. We also compared various types of scalar verification measures among four subsets of data, whose long-term event frequencies were almost the same. The forecast skill for 2003-2005 seemed to be the best among the four subsets; however, the better forecast skill for 2003-2005 seemed to be due to the persistence of solar activity. The judgment skill seemed to be the best during 2011-2015. Finally, we proposed the use of the judgment skill score to assess the judgment skill of an operational solar flare forecast and the verification strategy for a dichotomous and a multi-categorical operational solar flare forecast. \par

\section*{Appendix 1}
We briefly introduce the definitions of the scalar verification measures for the dichotomous forecast used in this study. Table \ref{tbl:dichoto_table} is a contingency table for the dichotomous forecast.

\begin{description}
\item[\parbox{6in}{Base rate (S).}] $$S=\frac{a+c}{a+b+c+d}=p(o)$$
\item[\parbox{6in}{Probability of detection (POD). Measure of discrimination.}] $$POD=\frac{a}{a+c}=p(f\ |\ o)$$
\item[\parbox{6in}{Probability of false detection (POFD). Measure of discrimination.}] $$POFD=\frac{b}{b+d}=p(f\ |\ \overline{o})$$
\item[\parbox{6in}{False alarm ratio (FAR). Measure of reliability.}] $$FAR=\frac{b}{a+b}=\frac{(1-S)POFD}{S\cdot POD+(1-S)POFD}=p(\overline{o}\ |\ f)$$
\item[\parbox{6in}{Proportion correct (PC). Measure of accuracy.}] $$PC=\frac{a+d}{a+b+c+d}=S\cdot POD+(1-S)POFD=p(f,o)+p(\overline{f},\overline{o})$$
\item[\parbox{6in}{Critical success index (CSI), also known as the threat score. Measure of accuracy.}] $$CSI=\frac{a}{a+b+c}=\frac{S\cdot POD}{S+(1-S)POFD}=p\left(f,o\ \Big|\ \overline{\overline{f},\overline{o}}\right)$$
\item[\parbox{6in}{Frequency bias (FB). Measure of bias.}] $$FB=\frac{a+b}{a+c}=POD+\frac{1-S}{S}POFD=\frac{p(f)}{p(o)}$$
\item[\parbox{6in}{Equitable threat score (ETS), also known as the Gilbert skill score. Measure of skill.}] $$ETS=\frac{a-a_r}{a-a_r+b+c}=\frac{S(1-S)(POD-POFD)}{S(1-S\cdot POD)+(1-S)^2POFD}$$ \par
$$a_r=\frac{(a+b)(a+c)}{a+b+c+d}$$
\item[\parbox{6in}{Heidke skill score (HSS). Measure of skill.}] $$HSS=\frac{PC-PC_r}{1-PC_r}=\frac{2S(1-S)(POD-POFD)}{S+S(1-2S)POD+(1-S)(1-2S)POFD}$$\par
$$PC_r=\frac{(a+c)(a+b)+(b+d)(c+d)}{(a+b+c+d)^2}$$
\item[\parbox{6in}{Peirce skill score (PSS), also known as the true skill statistic. Measure of skill.}] $$PSS=\frac{PC-PC_r}{1-PC_c}=POD-POFD$$ \par
$$PC_r=\frac{(a+c)(a+b)+(b+d)(c+d)}{(a+b+c+d)^2}\quad PC_c=\frac{(a+c)^2+(b+d)^2}{(a+b+c+d)^2}$$
\item[\parbox{6in}{Odds ratio skill score (ORSS). Measure of association.}] $$ORSS=\frac{ad-bc}{ad+bc}=\frac{POD-POFD}{POD(1-POFD)+POFD(1-POD)}$$
\item[\parbox{6in}{Symmetric extremal dependence index (SEDI). Measure of performance of extreme event. This measure is undefined when any element in contingency table is zero.}] $$SEDI=\frac{\log [POFD(1-POD)]-\log [POD(1-POFD)]}{\log [POFD(1-POD)]+\log [POD(1-POFD)]}$$
\end{description}

\section*{Appendix 2}
We briefly introduce the definition of the Gandin-Murphy-Gerrity score (GMGS) used in this study. A content of this Appendix is following Gandin \& Murphy (1992) and Gerrity (1992). \par

Define an $N$-categorical contingency table ${\bf P}$ with an element $p_{ij}$, which is a relative frequency of an observation falling category $i$ and a forecast falling category $j$. The GMGS is calculated as
\begin{equation}
	GMGS={\rm Tr}\ ({\bf S^{T}}\cdot {\bf P}),
\label{eq:gmgs}
\end{equation}
where ${\bf S}$ is an $N$-rank scoring matrix with an element $s_{ij}$. The $s_{ij}$ is determined for the GMGS satisfying equitability condition. The condition is written as follow.
\begin{equation}
	\sum_{i=1}^{N} s_{ji} p_i=0; \ j=1,\cdots,N,
\label{eq:gmgs_noskill}
\end{equation}
\begin{equation}
	\sum_{i=1}^{N} s_{ii} p_i=1,
\label{eq:gmgs_perfect}
\end{equation}
where $p_i$ is relative observation frequency of category $i$ defined as
\begin{equation}
	p_{i}=\sum_{j=1}^{N} p_{ij}.
\label{eq:gmgs_defpi}
\end{equation}
Equation (\ref{eq:gmgs_noskill}) means that scores of the GMGS for the forecast with always category $i$ vanish. Equation (\ref{eq:gmgs_perfect}) means that score of the GMGS for perfect forecast is one. Symmetry of scoring matrix ($s_{ij}=s_{ji}$) is also imposed. \par
Gerrity (1992) found a closed form scoring matrix satisfying the all conditions described above. The scoring matrix is defined as
\begin{equation}
	a_i=\frac{1-\sum_{k=1}^{i}p_k}{\sum_{k=1}^{i}p_k}; \ i=1,\cdots, N,
\end{equation}
\begin{equation}
	s_{ii}=\frac{1}{N-1}\left[\sum_{k=1}^{i-1}a_k^{-1}+\sum_{k=i}^{N-1}a_k\right]; \ i=1,\cdots, N,
\end{equation}
\begin{equation}
	s_{ij}=\frac{1}{N-1}\left[\sum_{k=1}^{i-1}a_k^{-1}+\sum_{k=i}^{j-1}(-1)+\sum_{k=j}^{N-1}a_k\right]; \ 1\le i<j\le N,
\end{equation}
\begin{equation}
	s_{ji}=s_{ij}.
\end{equation}
As we can recognize from the definition that the elements of scoring matrix are determined by only the observation frequency, so they do not depend on the forecast frequency. \par

The GMGS has a score of zero for unskillful forecast and one for perfect forecast. The GMGS for $N$-categorical contingency table is mathematically equals to an arithmetic mean of ($N-1$)-number PSSs, which are calculated from dichotomous contingency tables collapsed with threshold $k$ ($k=1,\cdots,N-1$). Therefore, the GMGS reaches the PSS for a case of dichotomous forecast (i.e. $N=2$). \par

\section*{Appendix 3}
We briefly show that a ranked probability score (RPS) applied to a multi-categorical deterministic forecast does not satisfy equitability. \par
An RPS for $N$-categorical probabilistic forecast is defined as 
\begin{equation}
	RPS=E\left(\frac{1}{N-1}\sum_{n=1}^{N-1}\left(F_n-O_n\right)^2\right),
\label{eq:RPSdef}
\end{equation}
where $F_n$ and $O_n$ are cumulative probabilities at category $n$ of forecast and observation, respectively (e.g., Jolliffe \& Stephenson 2003). The expectation value $E(\cdots)$ is calculated for all forecast-observation pairs. When an observation falls category $i$, the $O_n$ is zero and one for $n < i$ and $n \ge i$, respectively. If the forecast probabilities are set as one for category $j$ and zero for others (meaning deterministic forecast for category $j$), the $F_n$ is zero and one for $n < j$ and $n \ge j$, respectively. Therefore, the RPS applied to the deterministic forecasts can be written as 
\begin{equation}
	RPS=E\left(\frac{|j-i|}{N-1}\right).
\label{eq:RPSdeterministic}
\end{equation}
When an $N$-categorical contingency table ${\bf P}$ with an element $p_{ij}$, which is a relative frequency of an observation falling category $i$ and a forecast falling category $j$, is constructed from all forecast-observation pairs, equation (\ref{eq:RPSdeterministic}) can be rewritten as 
\begin{equation}
	RPS=\sum_{j=1}^{N} \sum_{i=1}^{N} s_{ji}p_{ij} = {\rm Tr}\ ({\bf S^{T}}\cdot {\bf P}),
\label{eq:RPStable}
\end{equation}
where $s_{ij}=|j-i|/(N-1)$. Therefore, the RPS applied to the deterministic forecast can mathematically be expressed by similar form of the GMGS. \par

Equitability requires a condition of equation (\ref{eq:gmgs_noskill}) to be satisfied (more precisely, the condition does not require zero for right hand side of the equation (\ref{eq:gmgs_noskill}), but a certain constant $c$ to be enough). The condition can be written by using the $s_{ij}$ determined for the RPS as 
\begin{equation}
	\frac{1}{N-1}\sum_{i=1}^{N} |j-i|p_{i}=c; \ j=1,\cdots,N,
\label{eq:RPScondition}
\end{equation}
where $p_i$ is relative observation frequency of category $i$. It is obvious that satisfying equation (\ref{eq:RPScondition}) for any $p_i$ is impossible. Therefore, the RPS applied to the deterministic forecast cannot satisfy the condition of equation (\ref{eq:gmgs_noskill}), and does not satisfy equitability. \par

\begin{acknowledgements}
We would like to thank SWPC/NOAA for compiling the GOES X-ray flare events list. We also would like to thank anonymous referees for useful comments to improve the manuscript. The editor thanks David Jackson and an anonymous referee for their assistance in evaluating this paper.
\end{acknowledgements}






\newpage

\begin{table}
	\begin{tabular}{|l|l|l|l|}
		\hline
		Code & Word & Flare class & 1-8\AA\ X-ray flux: $F$ (Wm$^{-2}$) \\ \hline\hline
		0 & Quiet & B or lower & $F <$ 10$^{-6}$ \\ \hline
		1 & Eruptive & C & 10$^{-6}$ $\leq F <$ 10$^{-5}$ \\ \hline
		2 & Active & M & 10$^{-5}$ $\leq F <$ 10$^{-4}$\\ \hline
		3 & Major flare & X & 10$^{-4}$ $\leq F$ \\
		\hline
	\end{tabular}
\caption{Definition of Regional Warning Center Japan's operational solar flare forecast.}
\label{tbl:flare_class}
\end{table}

\begin{table}
	\includegraphics[width=8cm]{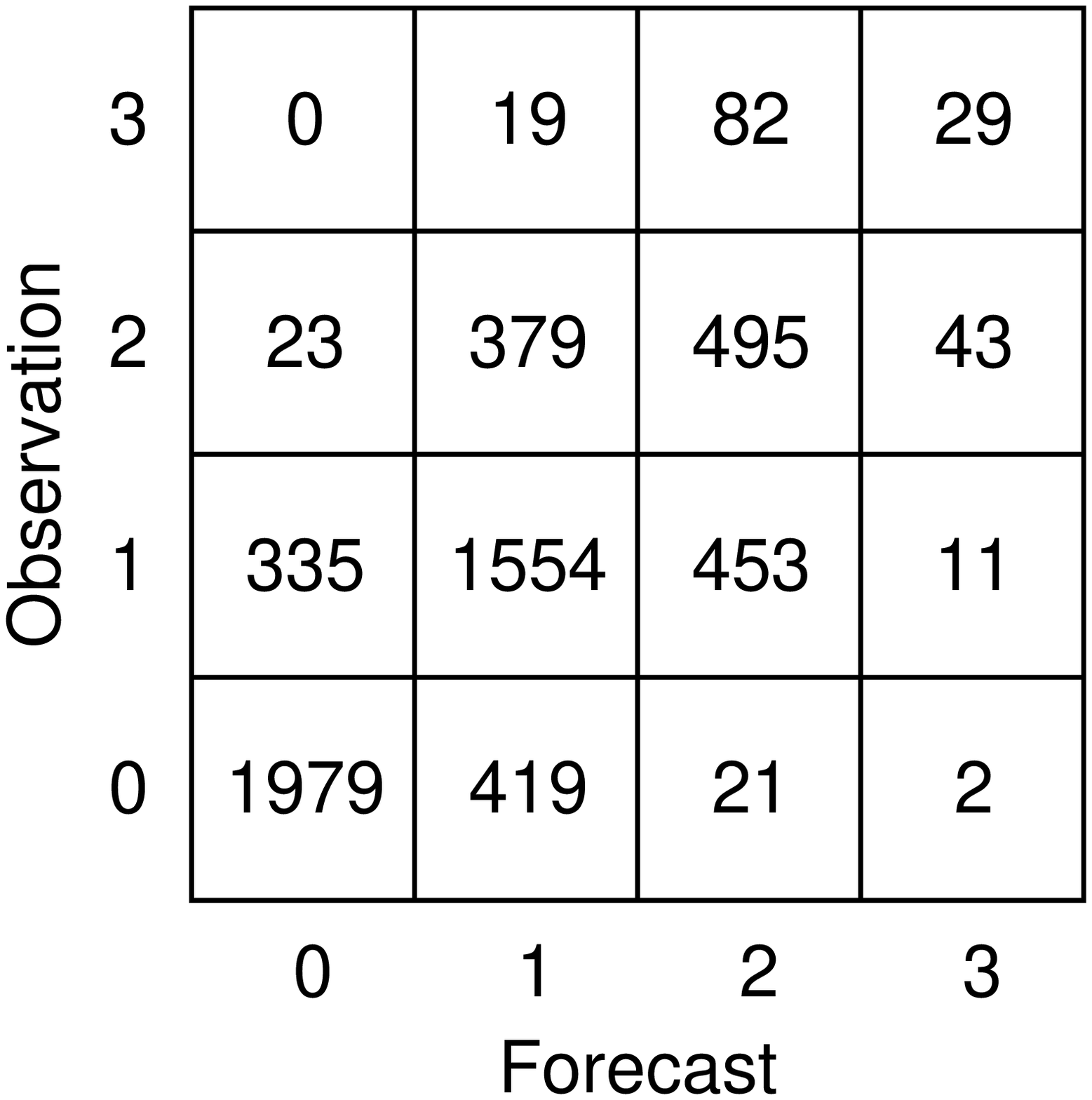}
	\caption{Four-categorical contingency table for the RWCJ forecast from 2000 to 2015.}
	\label{tbl:ct}
\end{table}

\begin{table}
	\begin{tabular}{|l|l|l|}
		\hline
		Mandatory & Highly recommended & Recommended \\ \hline\hline
		Hits & Frequency bias (FB) & Probability of false detection (POFD) \\
		Misses & Proportion correct (PC) & Critical success index (CSI) \\
		False alarms & Probability of detection (POD) & Peirce skill score (PSS) \\
		Correct rejections & False alarm ratio (FAR) & Heidke skill score (HSS) \\
		& Equitable threat score (ETS) & Odds ratio (OR) \\
		& & Odds ratio skill score (ORSS) \\
		& & Extremal dependence index (EDI) \\ \hline
	\end{tabular}
\caption{Recommended verification measures for deterministic forecasts of rare events (WMO 2014).}
\label{tbl:WMO}
\end{table}

\begin{table}
	\begin{tabular}{|l|r|r|c|}
		\hline
		Verification measures (attribute) & \multicolumn{1}{c|}{M $\leq$} & \multicolumn{1}{c|}{X $\leq$} & Perfect score \\ \hline\hline
		Hits & \multicolumn{1}{c|}{649} & \multicolumn{1}{c|}{29} & - \\
		Misses & \multicolumn{1}{c|}{421} & \multicolumn{1}{c|}{101} & - \\
		False alarms & \multicolumn{1}{c|}{487} & \multicolumn{1}{c|}{56} & - \\
		Correct rejections & \multicolumn{1}{c|}{4287} & \multicolumn{1}{c|}{5658} & - \\ \hdashline
		FB (bias) & 1.06 [1.01, 1.12] & 0.654 [0.519, 0.824] & 1 \\
		PC (accuracy) & 0.845 [0.835, 0.854] & 0.973 [0.969, 0.977] & 1 \\
		POD (discrimination) & 0.607 [0.577, 0.635] & 0.223 [0.157, 0.302] & 1 \\
		FAR (reliability) & 0.429 [0.400, 0.458] & 0.659 [0.551, 0.756] & 0 \\
		ETS (skill) & 0.327 [0.303, 0.353] & 0.147 [0.101, 0.205] & 1 \\ \hdashline
		POFD (discrimination) & 0.102 [0.0938, 0.111] & 0.009.80 [0.00752, 0.0126] & 0 \\
		CSI (accuracy) & 0.417 [0.393, 0.442] & 0.156 [0.109, 0.214] & 1 \\
		HSS (skill) & 0.493 [0.465, 0.522] & 0.257 [0.183, 0.340] & 1 \\
		PSS (skill) & 0.505 [0.474, 0.535] & 0.213 [0.147, 0.292] & 1 \\
		ORSS (association) & 0.863 [0.842, 0.882] & 0.933 [0.891, 0.959] & 1 \\
		SEDI (extreme) & 0.682 [0.652, 0.711] & 0.527 [0.439, 0.611] & asymptotic to 1 \\ \hdashline
		GMGS (skill) & \multicolumn{2}{c|}{0.477 [0.451, 0.506]} & 1 \\
		\hline
	\end{tabular}
\caption{Estimated scores of verification measures for events defined as larger than or equal to the M- and X-class flares.}
\label{tbl:verification_measure}
\end{table}

\begin{table}
	\begin{tabular}{rr|ll}
		\hline
		& & \multicolumn{1}{r}{Observed} & \\ 
		& & Yes & No \\ \hline
		\multirow{2}{*}{Forecast} & Yes & a (Hits) & b (False alarms) \\
		& No & c (Misses) & d (Correct rejections) \\
		\hline
	\end{tabular}
\caption{Contingency table for dichotomous forecast.}
\label{tbl:dichoto_table}
\end{table}

\begin{table}
	\begin{tabular}{|l|r|}
		\hline
		Verification measures (attribute) & \multicolumn{1}{c|}{Scores} \\ \hline\hline
		  PC$_{\mathrm m}$ (accuracy)  & 0.694 [0.683, 0.707] \\
									\hline
		  CC (association) 				& 0.717 [0.703, 0.730] \\
									\hline
		  GMGS (forecast skill) 			& 0.477 [0.451, 0.506] \\
									\hline
		  JS$_{\mathrm m}$ (judgment skill) & 0.0624 [0.00331, 0.116] \\
		\hline
	\end{tabular}
\caption{Estimated scores of multi-categorical verification measures for the overall data of RWCJ forecast.}
\label{tbl:multi_score}
\end{table}

\end{document}